\documentclass[twocolumn,superscriptaddress,showpacs,prl]{revtex4}    %
\usepackage{graphicx,amsmath,mathrsfs}
\usepackage{amssymb}
\usepackage{amsthm,multirow}
\usepackage{bm}
\usepackage{subfigure,xcolor}   

\usepackage{color}

\def\bc{\begin{center}}
\def\ec{\end{center}}

\newcommand{\bs}[1]{\boldsymbol{#1}}

\newcommand{\nn}{\nonumber}

\newcommand{\up}{\uparrow}
\newcommand{\dw}{\downarrow}
\newcommand{\pd}{{\phantom{\dagger}}}

\def\ie{\emph{i.e.},\ }

\begin{document}
\title{Density wave instabilities and surface state evolution in interacting Weyl semimetals}
\author{Manuel Laubach}
\affiliation{Institut f\"ur Theoretische Physik, Technische Universit\"at Dresden, 01062 Dresden, Germany}
\author{Christian Platt}
\affiliation{Department of Physics, Stanford University, Stanford, CA 94305, USA}
\author{Ronny\,Thomale}
\affiliation{Institut f\"ur Theoretische Physik, Universit\"at W\"urzburg, 97074 W\"urzburg, Germany}
\author{Titus Neupert}
\affiliation{Physik-Institut, Universit\"at Z\"urich, Wintherthurerstrasse 190, CH-8057 Z\"urich, Switzerland}
\author{Stephan\,Rachel}
\affiliation{Institut f\"ur Theoretische Physik, Technische Universit\"at Dresden, 01062 Dresden, Germany}
\affiliation{Department of Physics, Princeton University, Princeton, New Jersey 08544, USA}

\begin{abstract}
We investigate the interplay of many-body and band structure effects of interacting Weyl semimetals (WSM). 
Attractive and repulsive Hubbard interactions are studied within a model for a time-reversal-breaking WSM   with tetragonal symmetry, where we can approach the limit of weakly coupled planes and coupled chains by varying the hopping amplitudes.
Using a slab geometry, we employ the variational cluster approach to describe the evolution of WSM Fermi arc surface states as a function of interaction strength. 
We find spin and charge density wave instabilities which can gap out Weyl nodes. We identify scenarios where the bulk Weyl nodes are gapped while the Fermi arcs still persist, hence realizing a quantum anomalous Hall state.
\end{abstract}

\pacs{71.20.Gj, 71.10.Fd, 71.45.Lr, 03.65.Vf}


\maketitle


\paragraph{Introduction.}
Understanding the interplay of strong correlations and topology is a challenging task of contemporary condensed matter research that holds the promise of uncovering new physical phenomena. Established experimental and theoretical efforts on studying topological insulators have recently been complemented by analyzing topological metals and semimetals, 
where interactions naturally play a more important role\,\cite{nielsen-83plb389,Volovik03,murakami07njp356,wan-11prb205101,burkov-11prl127205,turner-13arXiv:1301.0330}. 

Weyl semimetals (WSM) aspire to claim as much attention and potential relevance as the topological insulator paradigm: many experimental material proposals and realizations have been reported\,\cite{huang-15nc7373,xu-15s613,lu-15s622,soluyanov-15n495,lv-15prx031013}, the first wave of transport experiments on WSM has revealed unconventional properties \,\cite{hosur-13crp857,burkov15jpcm113201,xiong-1503.08179,huang-15prx031023,zhang-16nc10735,arnold-16nc11615}, and even more exotic states of matter descendent from WSM have been formulated\,\cite{jian-15prl237001,morimoto-16sr19853}. In particular, while interaction effects such as band renormalization can be considered at the level of a single-particle analysis, WSM materials also display collective many-body effects such as superconductivity\,\cite{qi-16nc11038} or magnetism\,\cite{wang-16arXiv:1603.00479,arXiv:1604.02124,Shekhar-16arXiv:1604.01641,arXiv:1608.03404}, which can only be reconciled by including many-body instabilities into the low-energy description. 
There have been several previous analytical studies on interaction effects in WSMs\,\cite{wei-12prl196403,wang-13prb161107,maciejko-14prb035126,witczakKrempa-14prl136402,lai-15prb235131,jian-15prl237001,jian-15prb045121,bi-15prb241109,morimoto-16sr19853,wang-16arXiv:1604.05311}. In contrast, numerical attempts are rare\,\cite{witczakKrempa-14prl136402,go-12prl066401}: (i) within cluster perturbation theory bulk band renormalization effects were taken into account%
\,\cite{witczakKrempa-14prl136402} and (ii) within cluster dynamical mean field theory (CDMFT) a pyrochlore model has been studied where for strong interactions a Weyl semimetal phase is stabilized\,\cite{go-12prl066401}.


 In this paper, we aim to study more systematically the role of repulsive and attractive Hubbard interactions in a time-reversal-breaking WSM both on the bulk and on its surface states. The phenomenological reason for including attractive interactions is the possibly important role of phonon-mediated contributions to the electronic interaction profile. 
Among the possible many-body instabilities, 
we constrain our attention to spin and charge density wave instabilities, and leave out superconducting instabilities for the time being. 
The presence of time-reversal symmetry breaking and spin-orbit coupling tends to make the formation of a superconducting phase less preferential, and the strong interactions at commensurate filling likewise make density wave instabilities more competitive.

\begin{figure}[b!]
\centering
\includegraphics[scale=0.11]{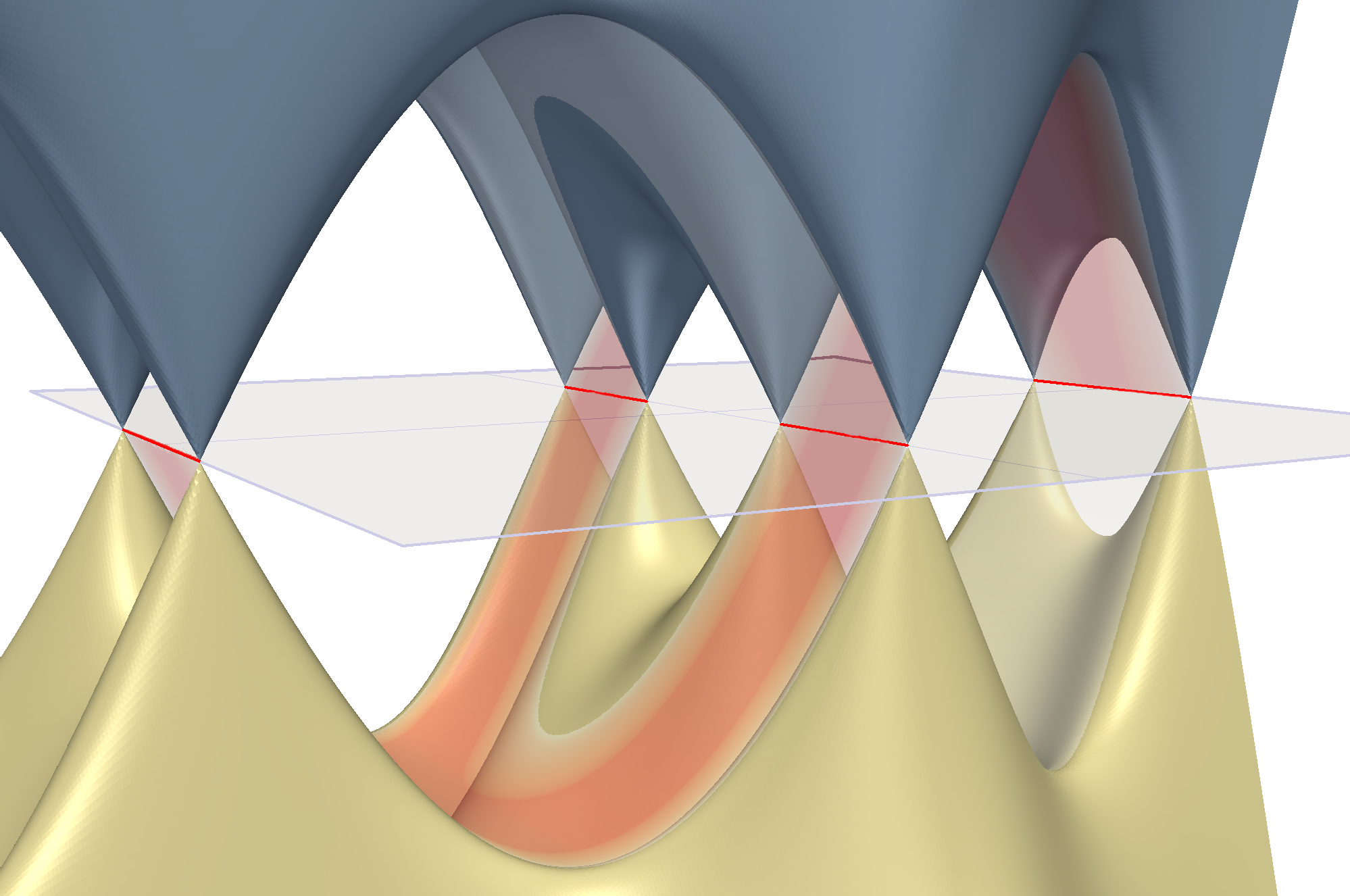}
\caption{Tight-binding bandstructure of the non-interacting WSM$_6$ phase ($m_x=-1$, $m_0=2$, $t_x=2$) in the surface BZ, which is shown in transparent grey with $(k_x, k_y) \in [-\pi,\pi]\times [-\pi,\pi]$. Surface state (light-red) including Fermi arc (red), the valence (gold) and conduction bulk bands (blue) with the Weyl nodes are shown. The surface state containing the three Fermi arcs evolves through the entire surface BZ.}
\label{fig:wsm-U0}
\end{figure}

\paragraph{Model.}
Realistic materials typically contain many bands which makes the study of strong correlation effects notoriously difficult. In order to gain a basic understanding it is instructive to consider simplified models with a minimal number of orbitals or bands\,\cite{delplace-12epl67004,yang-11prb075129,turner-13arXiv:1301.0330,witczakKrempa-14prl136402,chen-15arXiv:1507.00128}, respectively. In contrast to Dirac semimetals in 3D which feature degenerate bands, a Weyl semimetal is characterized by having no degeneracies in momentum space except at the nodal points. This requirement can only be fulfilled by breaking either inversion or time-reversal symmetry (or both). Material examples of the former category are TaAs\,\cite{PhysRevX.5.011029,huang-15nc7373,xu-15s613,lv-15prx031013}, (Mo/W)Te$_2$\,\cite{soluyanov-15n495} and of the latter  YbMnBi$_2$\,\cite{borisenko-15arXiv:1507.04847} as well as certain half-Heusler compounds\,\cite{wang-16arXiv:1603.00479,Shekhar-16arXiv:1604.01641}. 
We investigate a time-reversal breaking WSM where the  non-interacting model is defined on a simple tetragonal lattice via the Hamiltonian
\begin{eqnarray}\nn
\mathcal{H}_0 \!\!\!&=&\!\!\! \sum_{\bs{k}}\! c_{\bs{k}}^\dag \Big\{\! \big[ m_x \!-\! 2 t_x \cos{k_x} + m_0(2 \!-\!\cos{k_y} \!-\! \cos{k_z})\big] \sigma^x \\
&&\qquad\qquad + 2 t \sin{k_y} \sigma^y + 2 t \sin{k_z} \sigma^z \Big\} c_{\bs{k}}\pd\!\!,
\label{bandstructure}
\end{eqnarray}
which is a variant of the models considered in Refs.\,\cite{delplace-12epl67004,yang-11prb075129,turner-13arXiv:1301.0330,witczakKrempa-14prl136402,chen-15arXiv:1507.00128}.
Here $t_x$ denotes the hopping amplitude in the $x$ direction, $m_x$ and $m_0$ are further band structure parameters and $c_{\bs{k}} \equiv (c_{\bs{k},\up}, c_{\bs{k},\dw})$. We fix $t=1$ throughout this paper.
The Hamiltonian has $C_4$ rotational symmetry around the $x$ axes. Although this symmetry simplifies the analysis of the interacting case,  we do not expect any qualitative change of our results when relaxing this symmetry constraint.

The noninteracting phase diagram includes various WSM phases with two, four, six, and eight nodes which we label as WSM$_2$, WSM$_4$, WSM$_6$, and WSM$_8$, respectively\,\cite{wsm6}. For appropriately chosen surfaces, each pair of Weyl nodes is associated with one Fermi arc surface state in the corresponding surface Brillouin zone (BZ)\,\cite{footnote1}.
Due to the symmetry of the Hamiltonian, the positions of the Weyl-nodes in momentum space are unrestricted along the $k_x$-direction and are tied to the axes $(k_y,k_z) \in \left\{ (0,0), (\pi,\pi), (0,\pi), (\pi,0)\right\}$. 
Since the latter two are related by the $C_4$ symmetry of $\mathcal{H}_0$ 
Weyl nodes along these lines appear pairwise at the same momentum $k_x$.
In addition to the four different WSM phases, we obtain a normal insulating and a 3D quantum anomalous Hall (QAH) insulating regime; the QAH phase is stabilized upon approaching the limit of weakly coupled layers identifying this phase as a stacked Chern insulator regime.

One of the striking features of WSMs is the presence of {\it Fermi arc} surface states; that is, Fermi surfaces  which do not form closed loops in the BZ but lines which start at certain points in the surface BZ and end at another. 
In order to visualize the Fermi arc surface states we use a slab geometry, \ie open (periodic) boundary conditions along the $z$ ($x$ and $y$) direction. In Fig.\,\ref{fig:wsm-U0} we show the non-interacting surface state together with the bulk bands of the WSM$_6$ phase in the surface BZ for $t_x=2.0$. 
The surface state is easily identified by having almost the full weight of the eigenstates localized at the top surface.
The Fermi arcs (\ie the surface state restricted to energy $\omega=0$) are highlighted as red lines being part of the surface state (in light-red) which spans over the entire BZ [see also Refs.\,\onlinecite{haldane14arXiv:1401.0529,fang-15arXiv:1512.01552}]. 

\begin{figure}[t!]
\centering
\includegraphics[scale=0.88]{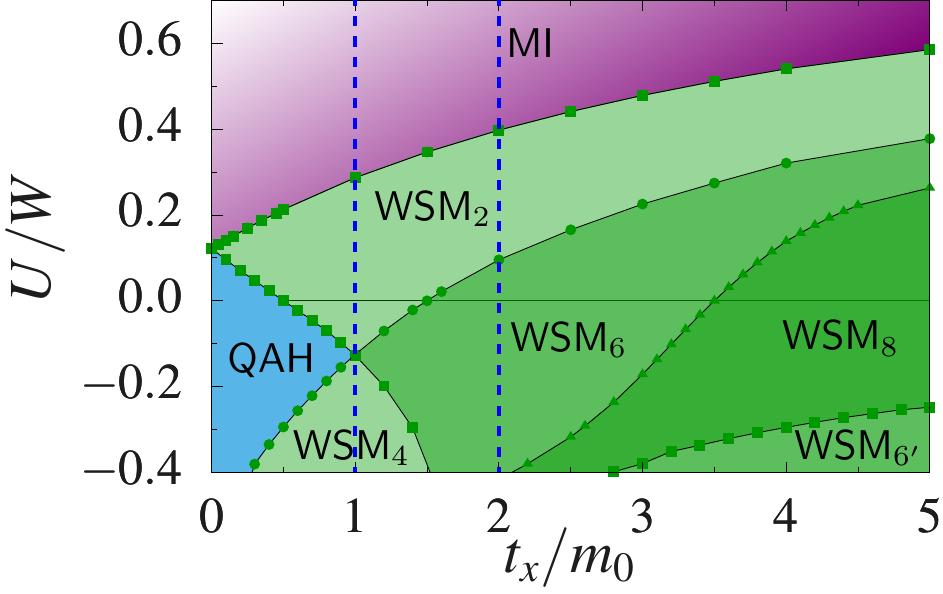}
\caption{``Paramagnetic'' interacting phasediagram (\ie ignoring many-body instabilitites) containing a QAH insulating state, various WSM phases which are distinguished by the number of Weyl nodes\,\cite{wsm6}, and a trivial Mott insulator. $W$ denotes the bandwidth.}
\label{fig:phasediagram}
\end{figure}

We reduce the parameter space of the bandstructure by fixing $m_x=-1$ and $m_0=2$ leaving $t_x \geq 0$ as the only free parameter allowing for a QAH, 
WSM$_2$, WSM$_6$, and WSM$_8$ phases [{\it cf.} Fig.\,\ref{fig:phasediagram} at $U=0$]. For $t_x\ll 1$ we approach the limit of weakly coupled planes while for $t_x \gg 1$ we obtain an effective 1D regime. In the following, we investigate the effect of local Hubbard interactions,
\begin{equation}\label{ham}
\mathcal{H}_I=U \sum_i n_{i\up} n_{i\dw}\ 
\end{equation}
considering both the repulsive ($U>0$) and the attractive ($U<0$) regime. In order to account for interaction effects, we apply the {\it Variational Cluster Approach} (VCA)\,\cite{potthoff-03prl206402,potthoff03epjb335} which is a cousin of
CDMFT\,\cite{maier-05rmp1027}. Clusters containing $2\times 2\times 2=8$ sites are solved within exact diagonalization and then used to construct the interacting Green's function of the Hubbard model. While the self-energy is restricted to the momentum resolution of the cluster, the self-consistent and variational scheme of VCA provides the best-possible approximation for the true self-energy (with the finite size of the cluster being the only source of errors). Note that quantum fluctuations within the cluster are fully taken into account by the VCA.
 Symmetry breaking many-body instabilities such as charge and spin density waves are implemented by means of Weiss-fields (see below). 
 The VCA has been successfully applied  in several fields of strongly correlated elctrons\,\cite{Senechal2005,balzer-10prb144516,AichhornArrigoniPotthoffEtAl2006,Sahebsara2008,KyungTremblay2006,laubach-14prb165136,Hassan2013b}.
 The method and its foundations are discussed in detail in the supplement\,\cite{supp}.

\begin{figure}[t!]
\centering
\includegraphics[width=0.48\textwidth]{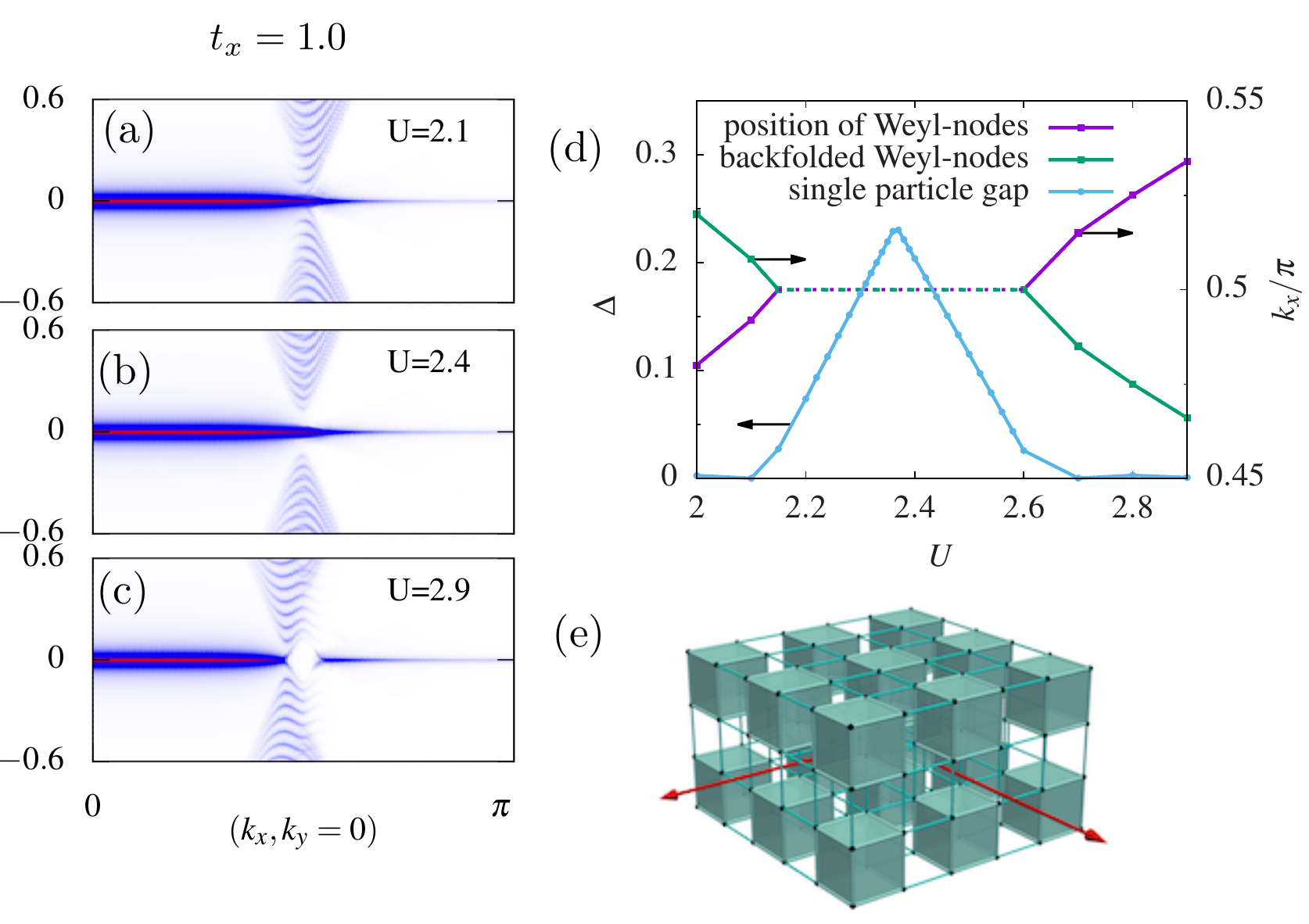}
\caption{
(a-c) Evolution of the interacting surface states at $t_x=1.0$ for repulsive interactions.
We show the high symmetry path $(k_x\!=\!0, k_y\!=\!0)\!\to\!(\pi,0)$. 
Note that the path $(0,0)\!\to\!(-\pi,0)$ is symmetric and therefore omitted.
We only select the spectral weight which is localized at the 
top surface guaranteeing the best resolution of the surface states. Note that the backfolded Fermi arc is plotted on a logarithmic scale to increase its visibility.
(b) Spectral function of the magnetic QAH phase with the gapped Weyl nodes. (a, c) Spectral functions of the magnetic WSM$_2$ phase.
(d) Position $k_x/\pi$ of the Weyl nodes and the single particle gap $\Delta$, 
identifying the QAH phase. (e) Schematic picture of the slab geometry with only two clusters (\ie four sites) along $z$. }
\label{fig:repulsive-tx1}
\end{figure}

Previously, we applied the VCA method in two spatial dimensions to investigate the helical edge states of topological insulators in the presence of interactions\,\cite{laubach-14prb165136,wu-12prb205102}. Most recently, we extended the VCA method to three-dimensional Hubbard models and studied antiferromagnetic and quantum paramagnetic phases\,\cite{laubach-15prb041106}.
Here we further extend it and construct a slab geometry within VCA for the first time. This development will allow us to track the surface states of the WSM and QAH phases even in the presence of interactions.
We generate a supercluster consisting of up to 40 clusters, 
\ie 80 lattice sites, which are stacked along the $z$-direction. 
This resulting supercluster is then used as unit cell for the subsequent VCA steps where we periodize the two remaining spatial directions, $x$ and $y$ [see Fig.\,\ref{fig:repulsive-tx1}\,(e) for an illustration].
In the following, we always show the single-particle spectral function $A(\omega, k_x, k_y)=-\frac{1}{\pi}{\rm Im}\{ G(\omega, k_x, k_y) \}$ of the surface BZ when we discuss interacting Fermi arcs.

In a first step, we investigate the ``paramagnetic'' solution of the {\it Weyl--Hubbard model} $\mathcal{H} = \mathcal{H}_0 + \mathcal{H}_I$, \ie we ignore any kind of many-body instabilities and only include the renormalization of $\mathcal{H}_0$ due to interactions. 
Weyl cones are still pinned to the high-symmetry lines parallel to the $k_x$ axis. The interaction can change the position of the cones and annihilate them pairwise along these lines.
For the WSM$_2$ phase, similar findings have been reported previously\,\cite{witczakKrempa-14prl136402}.
Depending on the sign of $U$ and  the value of $t_x$, all the different WSM phases with different number of Weyl nodes are present as well as the QAH phase and a fully gapped Mott insulator (MI) phase for large repulsive $U$, see Fig.\,\ref{fig:phasediagram}.

The QAH state for $t_x <0.5 $ possesses a nontrivial band gap which is reflected by the presence of a surface state extending through the entire BZ. At zero energy, the surface Fermi surface stretches along $(k_x,k_y=0)$ in the surface BZ.
Moreover, the Hall conductivity is quantized as $\sigma^{yz}= \frac{1}{c}  e^2/h$\,({\it c.f.} Ref.\,\onlinecite{witczakKrempa-14prl136402}), where $c$ is the lattice constant in $z$ direction.

\paragraph{Spin and charge density wave instabilities.}
We examine the possibility of charge and spin density wave orders  as prototypical many body instabilities of a WSM for attractive and repulsive interactions, respectively. We  add the following 
Weiss fields to our numerical calculations:
\begin{align}
  \mathcal{H}'_{\rm CDW}&=  \mathcal{C} 
  \sum_i e^{i \mathbf{Qr_i}} (n_{i\up} + n_{i\dw})\ ,\\
  \mathcal{H}'_{\rm SDW}&=  \mathcal{S} 
  \sum_i e^{i \mathbf{Qr_i}}  c^\dag_{i} \sigma^x c^\pd_i\ ,
  \label{eq:cdw+sdw}
\end{align}
where $\mathcal{C}$ and  $\mathcal{S}$ are variational parameters and $\bs{Q}$ takes the commensurate values $(\pi,0,0)$, $(0,\pi,\pi)$, and permutations as well as $(\pi,\pi,\pi)$.
Depending on $t_x$ and $U$ we observe stable solutions for several values of $\bs{Q}$ as discussed below. All these density wave instabilities have in common that they are associated with an increased unit cell in real space leading to back-folding of the corresponding BZ. In principle, Weyl nodes 
could be back-folded onto each other causing them to gap out if they possess opposite chirality. At the same time Fermi arc surface states will be backfolded onto each other; whether they cancel or persist (then realizing a QAH state) depends on several details which will be discussed in the paragraph {\it topological SDW phase}.

\begin{figure}[t!]
\centering
\includegraphics[width=0.46\textwidth]{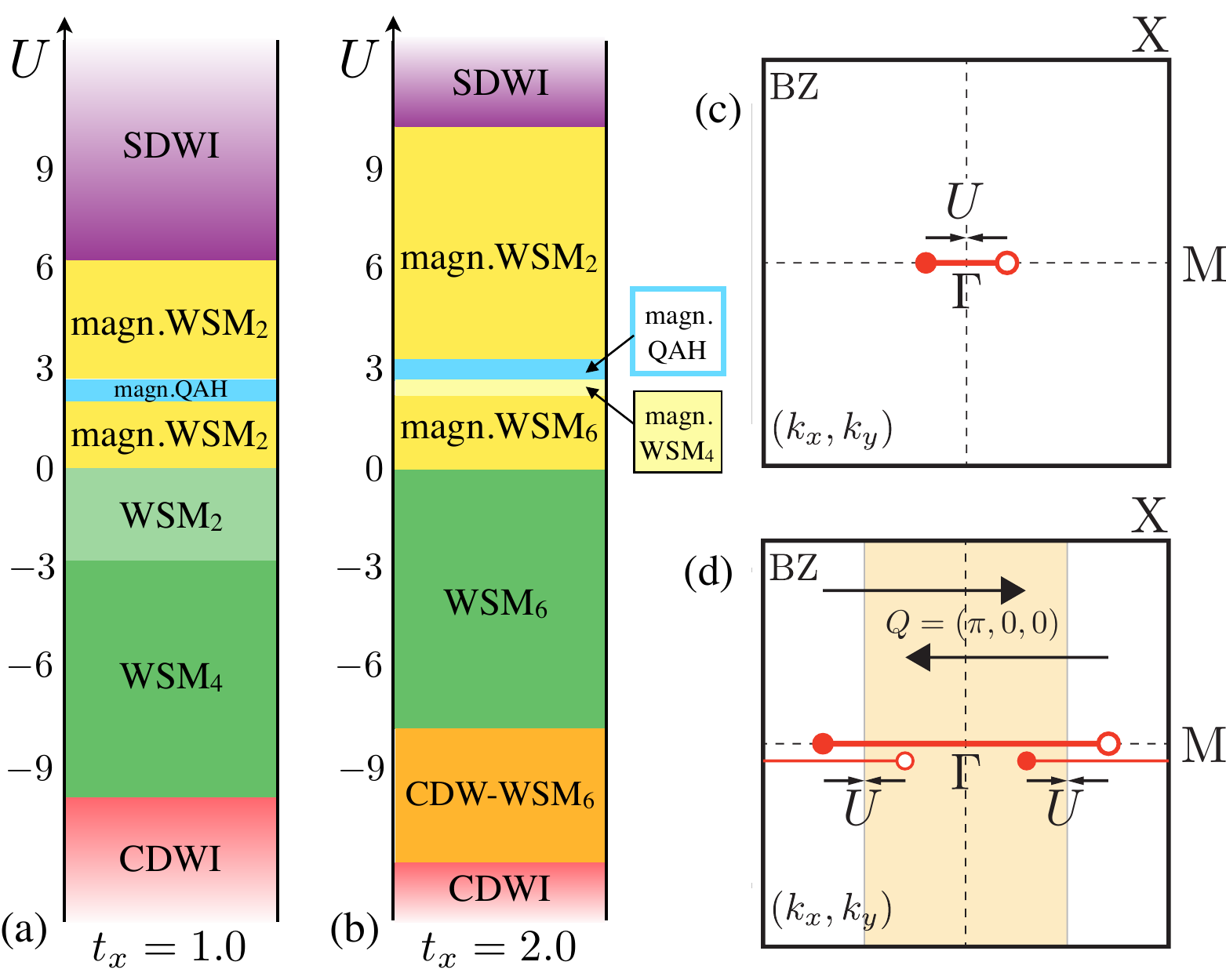} 
\caption{Interacting phase diagram including SDW and CDW orders for (a) $t_x=1$ and (b) $t_x=2$. 
Besides the normal semimetallic phases we also find them accompanied by SDW/CDW orders. Moreover, trivial insulating SDW and CDW phases appear as well as an antiferromagnetic QAH phase.
For discussion see main text. (c) Movement of two Weyl nodes towards each other influenced by the repulsive interaction leads to simultaneous annihilation of the Weyl nodes at $k_x=0$ and the Fermi arc. (d) Backfolding of the BZ can lead to doubling of the Fermi arc realizing a QAH state after the Weyl nodes annihilated at $k_x=\pi/2$. 
}
\label{fig:phasediagram_sdw+cdw}
\end{figure}

\paragraph{Interacting phase diagram.}
In Fig.\,\ref{fig:phasediagram_sdw+cdw}, we extend the previously discussed ``paramagnetic'' phase diagram in Fig.\,\ref{fig:phasediagram} to the full interacting phase diagram including many-body instabilities for representative cuts at $t_x=1$ and $t_x=2$.
Let us first consider the case $t_x=1$, see Fig.\,\ref{fig:phasediagram_sdw+cdw}\,(a). Since the non-interacting band structure explicitly breaks both time-reversal and spin rotation symmetry, antiferromagnetic order is stabilized for arbitrary small $U>0$. We find an ordering vector $\bs{Q}=(\pi,0,0)$ with magnetization pointing in the $x$ direction to be energetically favorable. That is, the magnetic BZ is backfolded along the $k_x$ direction.
We refer to such phases where a WSM coexists with a finite antiferromagnetic magnetization as ``magnetic WSM''. Note that such a coexistence  has also been discussed in 
Ref.\,\cite{wang-16arXiv:1603.00479,arXiv:1604.02124,Shekhar-16arXiv:1604.01641,arXiv:1608.03404}.

For $t_x=1$ and $U>0$, the non-interacting WSM$_2$ phase (present at $U=0$) becomes immediately antiferromagnetic 
(``magn.\ WSM$_2$''). Upon further increasing $U$, the Weyl nodes move on the axes parallel to $k_x$. 
At $U=2.12$, two Weyl nodes with opposite chirality meet at $k_x=\pi/2$ and annihilate. 
In addition, the backfolded Fermi arc is present, see Fig.\,\ref{fig:phasediagram_sdw+cdw}\,(d). At 
the transition point, the Fermi arc still extends through the entire BZ and becomes the surface state of the magnetic QAH phase. 
Note that the backfolded part of the surface state has a much weaker spectral weight [see surface spectral functions in Fig.\,\ref{fig:repulsive-tx1}] because its intensity is proportional to the SDW order parameter.
As the BZ is reduced compared to the QAH phase present for $U=0$ and small $t_x$, the Hall conductivity is halved, $\sigma^{yz} = \frac{1}{2c} e^2/h$,
 realizing half-QAH effect. At $U=2.63$ new Weyl nodes emerge and rebuild the magnetic WSM$_2$ phase. At $U=6.32$ the Weyl nodes meet again, the Fermi arc vanishes, and the system remains 
 in a topologically trivial insulating SDW phase (``SDWI'').

For $t_x=1$ and attractive $U<0$, 
the paramagnetic WSM phases remain stable until
at $U=-9.7$ finite CDW order sets in with $\bs{Q}=(\pi,\pi,\pi)$. The resulting backfolding of the BZ leads to pairwise annihilation of all Weyl nodes at $k_x=\,\pi/2$. 
In contrast to the repulsive case, all Weyl nodes and Fermi arcs are folded onto each other, such that a fully gapped phase appears (``CDWI''). We do not find a CDW-QAH phase as an CDW-analog of the magnetic QAH phase. In a fine-tuned situation, such a phase is likely to exist\,\cite{wang-16arXiv:1604.05311}.

The phase diagram for $t_x=2$ is shown in Fig.\,\ref{fig:phasediagram_sdw+cdw}\,(b). 
The WSM$_6$ phase present at $U=0$ [see Fig.\,\ref{fig:wsm-U0}] turns for finite $U>0$ immediately into a magnetic WSM$_6$ phase.
Upon further increasing $U$ we find the following sequence of phases: a magnetic WSM$_4$ phase (at $U=2.3$), a magnetic QAH phase (at $U=2.7$), a magnetic WSM$_2$ phase (at $U=3.15$), and a topologically trivial SDWI phase (at $U=10.34$).
For attractive $U$ the WSM$_6$ phase is stable down to $U=-7.8$ where a CDW with ordering vector $\bs{Q}=(0,\pi,\pi)$ appears which does not gap the system and therefore realizes a semimetallic CDW phase (``CDW-WSM''). Eventually at $U=-11.8$  a fully gapped CDW phase with order $\bs{Q}=(\pi,\pi,\pi)$ becomes energetically favorable.

\begin{figure}[t!]
\centering
\includegraphics[width=0.48\textwidth]{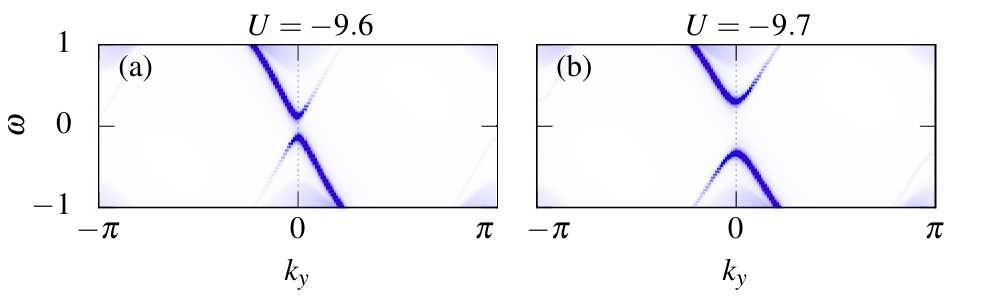} 
\caption{A representative cut through the surface states along $k_y$ (for fixed $k_x=3\pi/4$) is shown at the transition from WSM$_4$ into the CDWI phase ($t_x=1$). 
(a) Surface state with negative Fermi velocity and the backfolded surface state (possessing much weaker intensity) with positive velocity: they hybridize and acquire a small gap.
(b) Further increase of attractive interactions opens the gap further.
More details are delegated to the supplemental material\,\cite{supp}.}
\label{fig:spec-ortho}
\end{figure}

Note that the phase diagrams Figs.\,\ref{fig:phasediagram} and \ref{fig:phasediagram_sdw+cdw} are invariant under $t_x \to -t_x$. The results reported in this paper thus apply to a much broader parameter range. Moreover, note that repulsive but longer-ranged interactions represent another way of stabilizing CDW orders. We expect that a model with nearest-neighbor repulsions will feature  phases similar to those found for $U<0$ and possibly even a CDW-QAH phase.

\paragraph{Topological SDW phase.}
As mentioned before, backfolding due to an ordering vector $\bs{Q}$ maps Weyl nodes and Fermi arcs to other regions of the BZ where they might ``meet'' other nodes or arcs. In our model, repulsive (attractive) $U$ causes the Weyl nodes to move either towards the center (the edges) of the BZ or towards the edges (the center). 
Weyl nodes with opposite chirality gap out, as it is the case in the SDWI and CDWI phases. Weyl nodes with same chirality are unaffected as it is the case for the CDW-WSM or for the magnetic WSM phases. What dictates whether two Fermi arcs annihilate or not depends on the Fermi velocities of the their corresponding surface states. 
Surface states that possess Fermi velocities with opposite sign can hybridize and therefore gap out, which leads to vanishing Fermi arcs (see Fig.\,\ref{fig:spec-ortho} for an example); if the velocities have the same sign they  survive as the surface state of the topologically non-trivial QAH phase.

\paragraph{Conclusion.} 
We investigated the role of electron--electron interactions in Weyl semimetals and the possibility of spin and charge density wave orders. Using a slab geometry, we are able to track the interacting Fermi arc surface states.
We find spin and charge density wave instabilities which have the potential to gap out  Weyl nodes. We identify situations where all Weyl nodes are fully gapped, but the Fermi arcs are glued together forming a quantum anomalous Hall surface state due to backfolding associated with the spin density wave order.

{\it Acknowledgements.}~%
We thank D.\ A.\ Abanin and W.\ Witczak-Krempa for fruitful discussions. We acknowledge financial support
by the DFG through SFB 1143 (ML, SR) and SFB 1170 (RT),
by the ERC through the starting grant TOPOLECTRICS ERC-StG-Thomale-336012 (RT), 
and by the Helmholtz association through the virtual institute VI-521 (SR).
We thank the Center for Information Services and High Performance Computing (ZIH) at TU
Dresden for generous allocations of computer time.

\bibliographystyle{prsty}
\bibliography{wsm-vca}

\begin{thebibliography}{10}
\bibitem{footnote1}
It is worth mentioning that by choosing another surface the subsequent
  projection of the bandstructure onto the surface BZ changes as well; this
  might lead to situations where Fermi arcs either disappear or get doubled.
\end{thebibliography}

\end{document}